\documentclass{aastex631}

\usepackage{framed,multirow}

\usepackage{amssymb}
\usepackage{latexsym}

\usepackage{url}
\usepackage{xcolor}
\definecolor{newcolor}{rgb}{.8,.349,.1}
\usepackage{listings}
\usepackage{graphicx}
\usepackage{gensymb}
\usepackage{rotating}
\begin{document}

\title{A study of the kinematic and volumetric co-evolution of Earth-directed CMEs}

\author{Ashutosh Pattnaik}
\affiliation{Astronomical Observatory of the Jagiellonian University\\
Orla 171, 30-244 Kraków\\
Poland}

\author{Nat Gopalswamy}
\affiliation{NASA Goddard Space Flight Center\\
Maryland, USA}

\author{Ranadeep Sarkar}
\affiliation{University of Helsinki\\
Helsinki, Finland}

\author{Hong Xie}
\affiliation{NASA Goddard Space Flight Center\\
Maryland, USA}

\author{Sachiko Akiyama}
\affiliation{NASA Goddard Space Flight Center\\
Maryland, USA}

\author{Grzegorz  Micha$\l$ek}
\affiliation{Astronomical Observatory of the Jagiellonian University\\
Orla 171, 30-244 Kraków\\
Poland}
\begin{abstract}
While flare-associated CMEs generally show a strong association between flare X-ray flux and CME kinematics, their volumetric evolution and its link to both kinematics and flare activity remains less explored. In this study, we investigate the volumetric and kinematic co-evolution of ten Earth-directed, flare-associated CMEs using multi-viewpoint observations from STEREO-A, STEREO-B, and SOHO. We perform 3D reconstructions of the CME flux ropes with the Graduated Cylindrical Shell (GCS) model and derive their geometrical parameters. We find that the total CME volume follows a power-law dependence on the leading edge height, and that different structural components expand at different rates, with the ellipsoidal front expanding faster than the conical legs.
Furthermore, the volumetric evolution follows a multi-phase pattern: initial overexpansion, a gradual reduction in the expansion rate, and finally saturation at a higher heliocentric distance. This is similar to the well-established three-phase evolution of the CME kinematics. Notably, the second-order derivative of volume with time shows a strong temporal correlation with both CME acceleration and the GOES soft X-ray flux of the associated flare. This is the first study to report such a correspondence between volumetric evolution and flare timing, highlighting the role of flare energy release in governing CME expansion dynamics. Our findings motivate further studies into the coupling between magnetic reconnection and CME volumetric evolution in the corona.
\end{abstract}
\section{Introduction}
Coronal Mass Ejections (CMEs) are expulsions of large-scale plasma structures from the solar corona that encapsulate a magnetic flux rope structure within them \citep{Krall_2006,Vourlidas2013}. As CMEs propagate through the corona and into interplanetary space, their dynamics are governed primarily by the Lorentz force and the aerodynamic drag. These forces influence not only the kinematic evolution of CMEs but also contribute to their geometric deformation during propagation.

In the absence of significant external interactions, such as with preceding CMEs or high-speed solar wind streams (HSSs) emanating from coronal holes the dynamical evolution of a CME can be broadly divided into three phases. In the initial phase, the Lorentz force dominates, resulting in the impulsive acceleration of the CME. As the CME propagates farther through the corona and the Lorentz force diminishes, the drag force becomes increasingly significant, resulting in the deceleration of the CME. In the final phase, the CME reaches a state of dynamical equilibrium, propagating at a nearly constant speed that matches that of the ambient solar wind \citep{https://doi.org/10.1029/2001JA000177}.
In practice, however, the kinematic evolution of CMEs is often far more complex, primarily due to the intricate dynamical processes that arise during interactions with other CMEs or large-scale solar wind structures\citep{manchester2017}.
In this era, we have multi-viewpoint observations along the Sun–Earth line by coronagraphs and heliospheric imagers onboard missions such as Solar TErrestrial RElations Observatory (STEREO) and Solar Orbiter (SolO), alongside continuous solar monitoring from the L1 point by missions like SOlar and Heliospheric Observatory (SOHO), Solar Dynamics Observatory (SDO), and Aditya-L1. This observational capability has enabled the development of a variety of CME arrival forecasting approaches. These include empirical methods (e.g., \citealp{https://doi.org/10.1029/2001JA000177, Ravishankar2019, Pattnaik_2025}), analytical models such as the Drag-Based Model (DBM; \citealp{Vršnak2013}), semi-analytical techniques (e.g., \citealp{ELEVO-HI, ANTEATER-PARADE}), and MHD approaches (e.g., \citealp{Enlil, Euphoria}).

However, to advance our understanding of CME dynamics and improve predictions of their space weather impact, it is essential to investigate the geometrical evolution of CMEs. In particular, direct measurements of CME volume from remote sensing observations and the analysis of their volumetric expansion during propagation remain relatively unexplored in the current literature. Addressing this gap is likely to provide crucial insights into the physical processes that govern CME evolution and enhance forecasting capabilities.

\citet{Holzknecht} developed a method to determine CME volumes by fitting the Graduated Cylindrical Shell (GCS) model \citep{2009Thernisien, Thernisien_2011} to CME flux ropes. This method was later employed by \citet{Temmer-cmedensity} to investigate CME density evolution at heliocentric distances beyond $15~R_{\odot}$. However, studying CME dynamics closer to the Sun requires simultaneous, multi-viewpoint coronagraph observations covering the inner corona. Since October 2014, communication with STEREO-B has been lost, significantly complicating the 3D reconstruction of CMEs in this critical lower-corona regime. To overcome this observational gap, \citet{Majumdar_2022} introduced a novel technique combining observations from the ground-based MLSO/K-Cor coronagraph, effectively compensating for the lost inner-corona viewpoint previously provided by COR1 ($1.5~R_{\odot}-4~R_{\odot}$) onboard STEREO-B.

K-Cor has a field of view (FOV) of $1.05~R_{\odot}-3~R_{\odot}$, allowing it to probe even lower into the solar atmosphere than COR1(\citealp{Gopalswamy2012}, fig.15). \citet{Majumdar_2022} performed multi-viewpoint observations in the lower corona using K-Cor and COR1. K-Cor observations were replaced by data from the LASCO C2 coronagraph onboard SOHO whenever their FOV overlapped. For the outer corona, they utilized LASCO coronagraphs and COR2 onboard STEREO. The updated GCS method \citep{Majumdar_2022} could thus perform 3D reconstructions of CMEs continuously from the inner to the outer corona.
Their findings showed that the CME volumetric evolution follows different power laws in the inner corona and outer corona, implying that different mechanisms might be involved in CME volume expansion in these two regimes. They also demonstrated that different parts of CMEs follow different power laws in their evolution as well.
However, the mechanisms responsible for the shift in CME volumetric evolution during the transition from the inner to the outer corona remain poorly understood. In this manuscript, we will examine whether the volumetric evolution of CMEs follows a similar multi-phase mechanism as that of their kinematic evolution. We will also check if the solar flare associated with the CME affects its volumetric evolution.
We performed GCS reconstruction for a set of ten well-observed, Earth-directed CMEs using multi-vantage point observations from STEREO-A, STEREO-B, and SOHO. We utilized COR1 observations from both STEREO-A and STEREO-B, which provide coverage of the inner corona in the critical range of approximately $1.5$--$4.0~R_{\odot}$. And for the outer corona we utilised LASCO C2 coronagraph onboard SOHO and COR2 onboard STEREO-A and STEREO-B.

Section~\ref{sec:selection} outlines the criteria used for CME event selection. Section~\ref{CME-volume} and ~\ref{CME-evolution} describes the methodology adopted for determining CME volumes and studying their kinematic and volumetric evolution. The results are presented in Section~\ref{sec:results}, followed by a detailed summary and conclusions in Section~\ref{sec:conclusion}.

\section{Event Selection}
\label{sec:selection}
We examine the volumetric and kinematic evolution of a set of CMEs using the spatial parameters of the ejections and their propagation within the interplanetary medium. 
To ensure the quality and consistency of our research, we use the following  criteria to select the CMEs:

\begin{itemize}
    \item The CME must be associated with a soft X-ray flare.
    \item The CME should be non-interacting.
    \item The source region of the CME should be within \(\pm 45^{\circ}\) in longitude and latitude from the disc center of the Sun.
 \end{itemize}
 The first criterion enables us to establish a correlation between the parameters of solar eruptions and the expansion and propagation of CMEs. During the period of maximum solar activity we frequently observe collisions between CMEs during their propagation in the inerplanetary medium. The trajectories of interacting CMEs are unpredictable and require separate specific studies. Therefore, we excluded such CMEs. It is necessary to conduct a multi-viewpoint observation of the Earth-directed CMEs from the inner corona to the outer corona in order to accurately reconstruct the spatial parameters of the ejections. Therefore, it is necessary to observe the ejection simultaneously by the COR1 and COR2 coronagraphs on STEREO-A and B satellite and LASCO C2 and C3 coronagraphs onboard SOHO. Furthermore, the flux rope must be observed from the appropriate perspective to accurately reconstruct the CME volume. Consequently, to observe Earth-directed CMEs, which are generated in the center of the solar disk, it is preferred to utilize STEREO coronagraphs that are in quadrature with respect to the Earth.

From May 2010 to September 2011, the STEREO satellite maintained a quadrature configuration with respect to the Earth (\citealp{Gopalswamy_2013}). 
We were able to study the evolution of 10 well-observed CMEs from the inner corona to the outer corona by using these selection criteria. The details of flare-cme association was acquired from \href{https://cdaw.gsfc.nasa.gov/CME_list/halo/halo.html}{SOHO LASCO HALO CME catalog} \citep{2010SunGe...5....7G} and the \href{https://kauai.ccmc.gsfc.nasa.gov/DONKI/search/}{CCMC-DONKI} database.
  The table~\ref{CME_source_region} contains the list of selected CMEs and the information related to the associated flares. Table \ref{CME_source_region} includes the time of flare onset, the source location of the flare, the flare class, and the onset time of the associated CME.
 \begin{figure}[htbp]
    \centering
    \includegraphics[width=\textwidth]{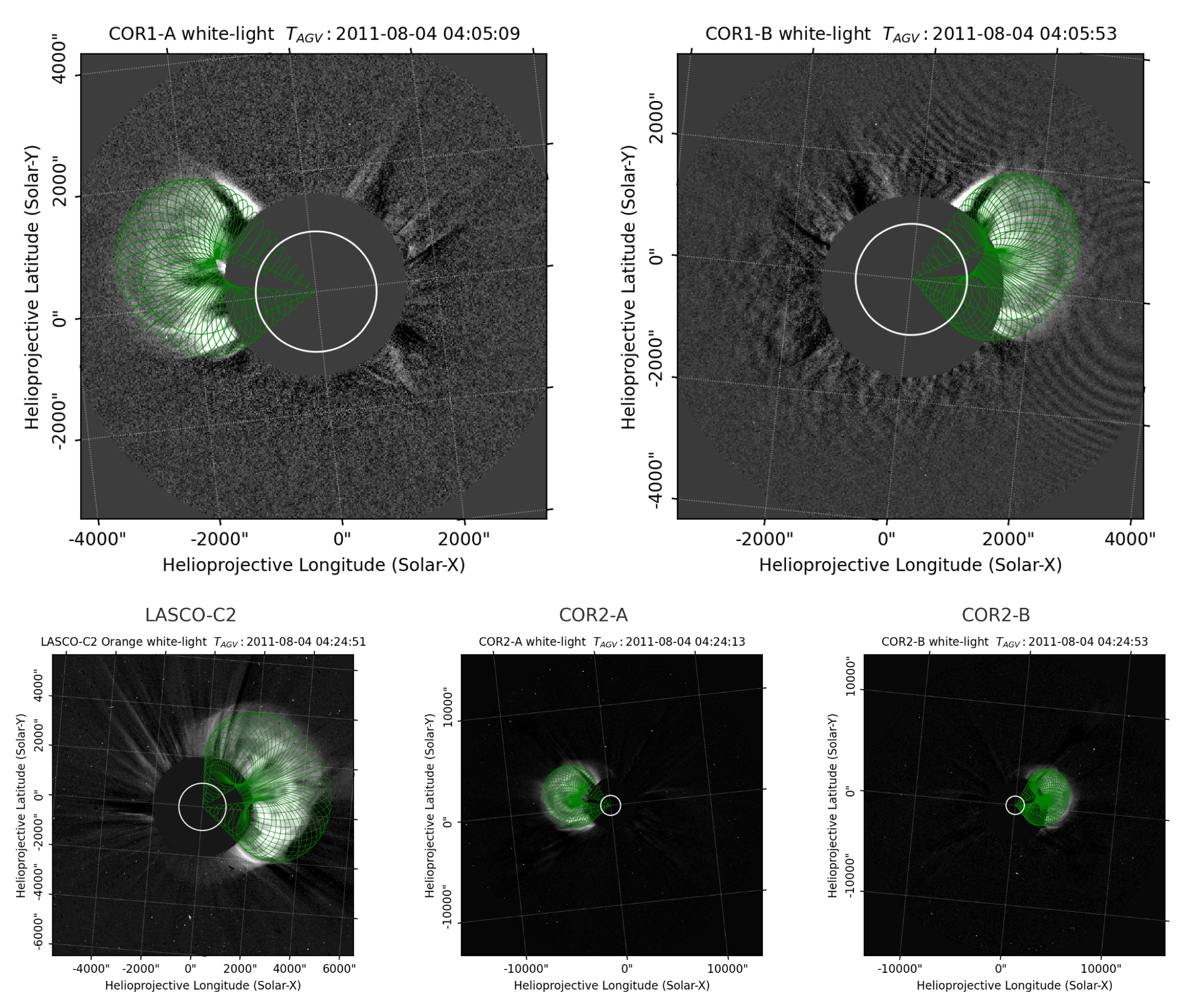}
    \caption{GCS reconstruction of the CME fluxrope for the event on 2011 August 4.}
    \label{GCS}
\end{figure}
Utilizing coronagraphic observations, we determined the volume and kinematic parameters of the investigated CMEs  during their propagation in the heliosphere. Subsequent sections delineate the methodologies employed to achieve this objective.
\section{Determination of CME volume}
\label{CME-volume}
By fitting the GCS model to each CME, we estimated the geometrical parameters of CME flux ropes, which are crucial for determining the volume of CMEs. We employed \texttt{PyThea} \citep{PyThea}, a Python-based program for CME three-dimensional reconstructions. Following the steps outlined in \cite{2009Thernisien}, the GCS grid was applied to each CME that was seen at almost the same time from the COR1-A and COR1-B coronagraph fields of view. The same method was used in the outer corona, where images from COR2-A, COR2-B, and LASCO-C2 were used to fit the GCS. Figure 1 shows the GCS reconstruction of the CME flux-rope for the August 4, 2011 event.

\begin{figure}[htbp]
    \centering
    \includegraphics[width=\textwidth]{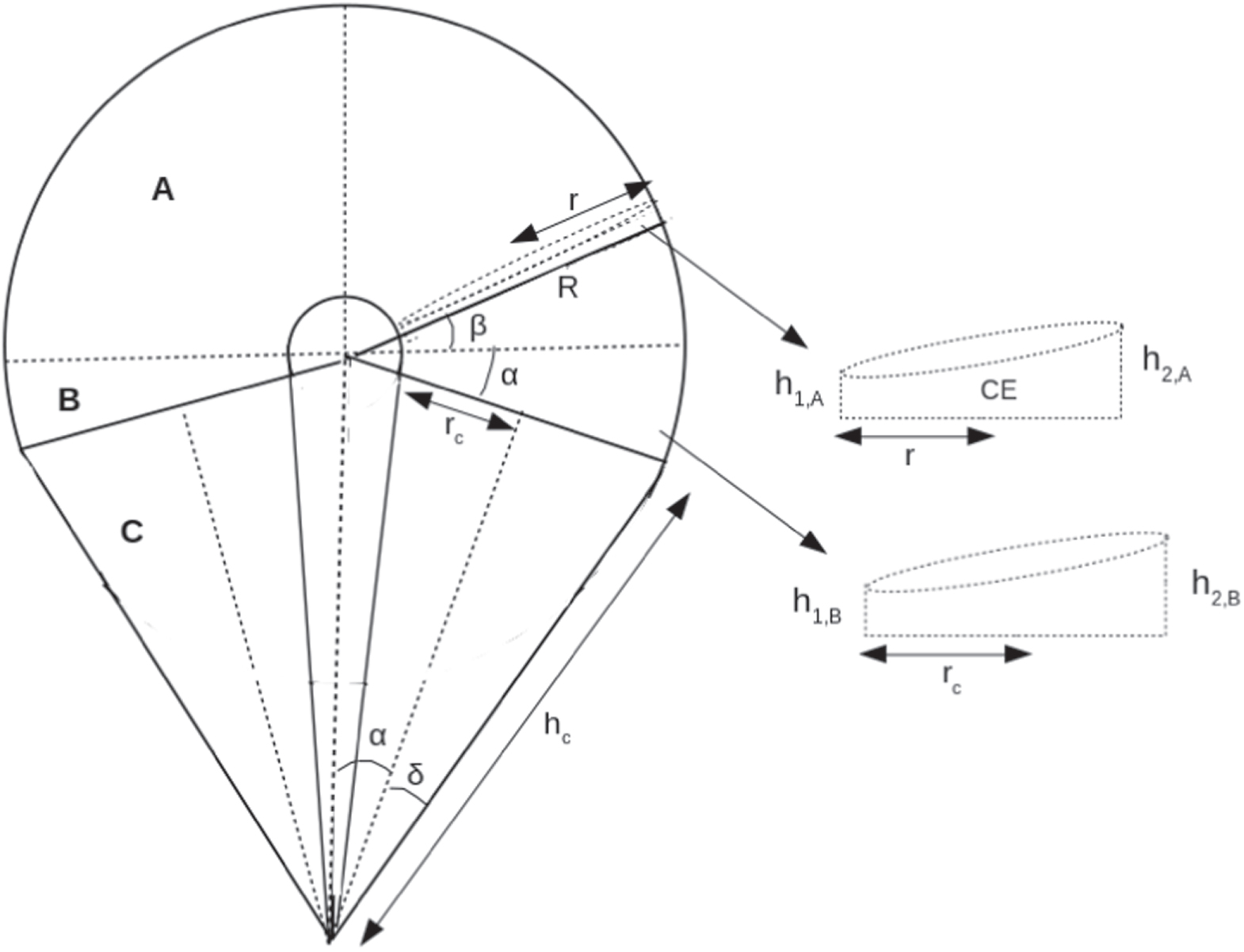}
    \caption{Distinct features of the CME fluxrope: the ellipsoidal front (part A), the assymetrical disc (part B) and the conical leg (part C). Acquired from \citet{Majumdar_2022}.}
    \label{fluxropevolume}
\end{figure}

This 3D reconstruction yields the following geometrical parameters of the flux rope: longitude ($\phi$), latitude ($\theta$), leading edge height ($h_{front}$), half-angle ($\alpha$), aspect ratio ($\kappa$), and tilt angle ($\gamma$) of the flux rope relative to the solar equator.

The fluxrope is first split into three parts: part A is the ellipsoidal front, part B is the asymmetric disc, and part C is the conical leg (see figure \ref{fluxropevolume}).
Following the instructions provided in \citet{Holzknecht} and \citet{Majumdar_2022} we can easily determine the volume of  (e.g., $Vol_a, Vol_b, Vol_c)$ and the total volume ($Vol,$) of the fluxrope using only three GCS parameters ($\alpha$, $\kappa$, and $h_{front}$).

\section{propagation and evolution of the CMEs}
\label{CME-evolution}

By computing the first and second time derivatives of the leading edge height, obtained from the GCS fitting of each CME, we derived the corresponding instantaneous speed and acceleration profiles. Similarly, the first and second time derivatives of the CME volume were calculated to characterize its volumetric evolution. Then, the evolution of these parameters in terms of their distance from the Sun and time were compared. As shown in the following sections.

Next, we looked at possible links between the CME evolution and the solar flare activity by comparing the GOES X-ray flux time series with the volumetric and kinematic profiles that we had made.
\section{Results}
\label{sec:results}

We plotted the CME volume as a function of the CME leading edge height and applied both a power-law and a quadratic polynomial fit to the data. To evaluate the goodness of the fit, we used the coefficient of determination (\(R^2\)) and the \(p\)-value. The \(R^2\) value quantifies the proportion of variance in the data that is explained by the model, with a value of 1 representing a perfect fit. The \(p\)-value, on the other hand, helps assess the statistical significance of the model by estimating the probability that the observed fit could have occurred by chance; lower values imply stronger evidence against the null hypothesis. In our case, the power-law fit consistently showed high \(R^2\) values and low \(p\)-values, suggesting that it better captures the relationship between CME volume and leading edge height.

We selected two representative CME events to illustrate our results. In section~\ref {volume}, we present the volumetric evolution of these events. Section~\ref{flare-cmekinematics} and ~\ref{flare-cmevolume} highlight our findings on the influence of associated solar flares on CME kinematic and volumetric evolution, respectively.
\begin{figure}[htbp]
    \centering
    \includegraphics[width=\textwidth]{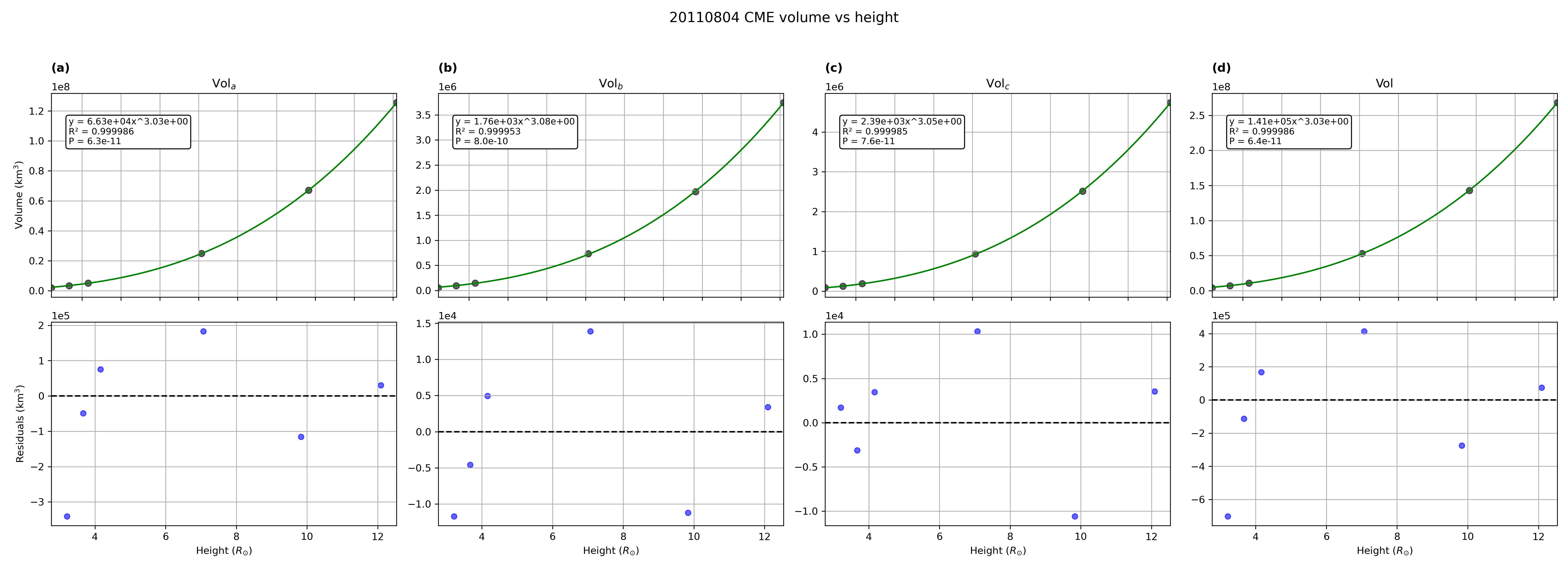}
    \caption{Power-law fits to each volume component (e.g., $Vol_a, Vol_b, Vol_c)$ and the total volume, $Vol,$ as a function of height for the CME on 2011 August 4. The top row contains the volume-height data points, represented by black dots, and the power law fit is represented by the green line. The bottom row shows the corresponding residuals.}
    \label{powerlaw_20110804}
\end{figure}

\begin{figure}[htbp]
    \centering
    \includegraphics[width=\textwidth]{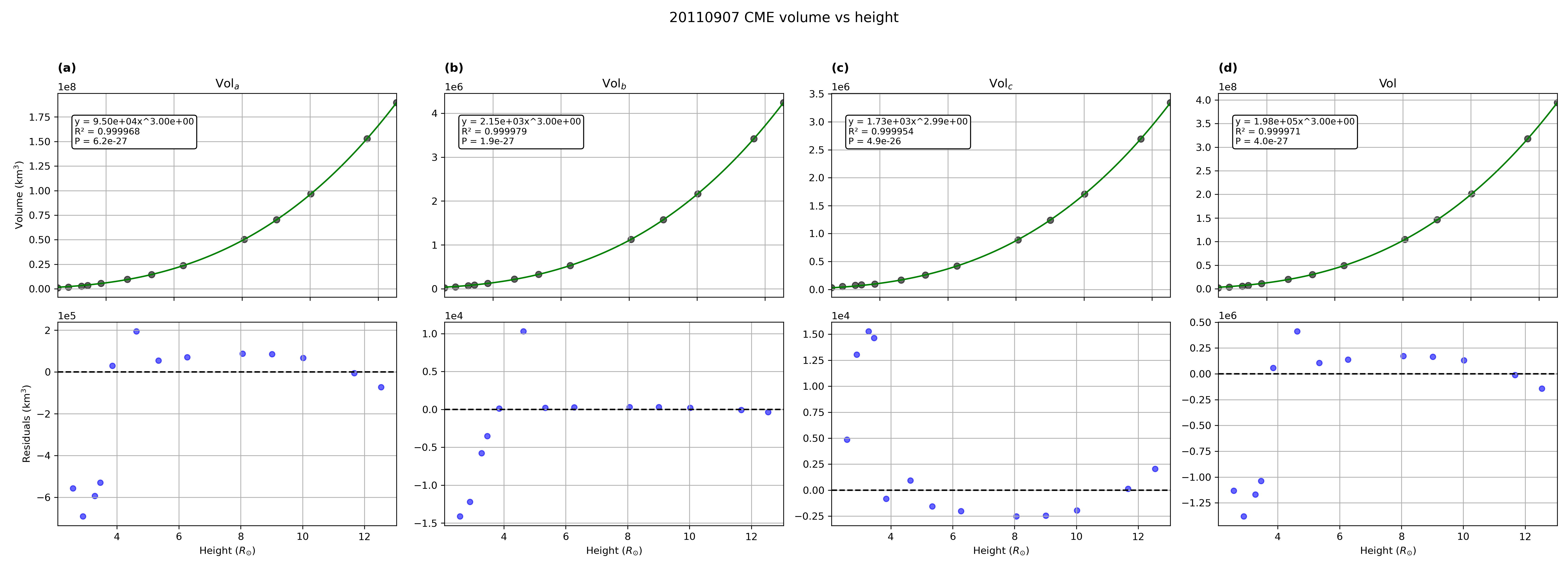}
    \caption{Power-law fits to each volume component (e.g., $Vol_a, Vol_b, Vol_c)$ and the total volume, $Vol,$ as a function of height for the CME on 2011 September 7. The top row contains the volume-height data points, represented by black dots, and the power law fit is represented by the green line. The bottom row shows the corresponding residuals.}
    \label{powerlaw_20110907}
\end{figure}

\subsection{Volumetric evolution of the CMEs}
\label{volume}
\begin{figure}[htbp]
    \centering
    \includegraphics[width=\textwidth]{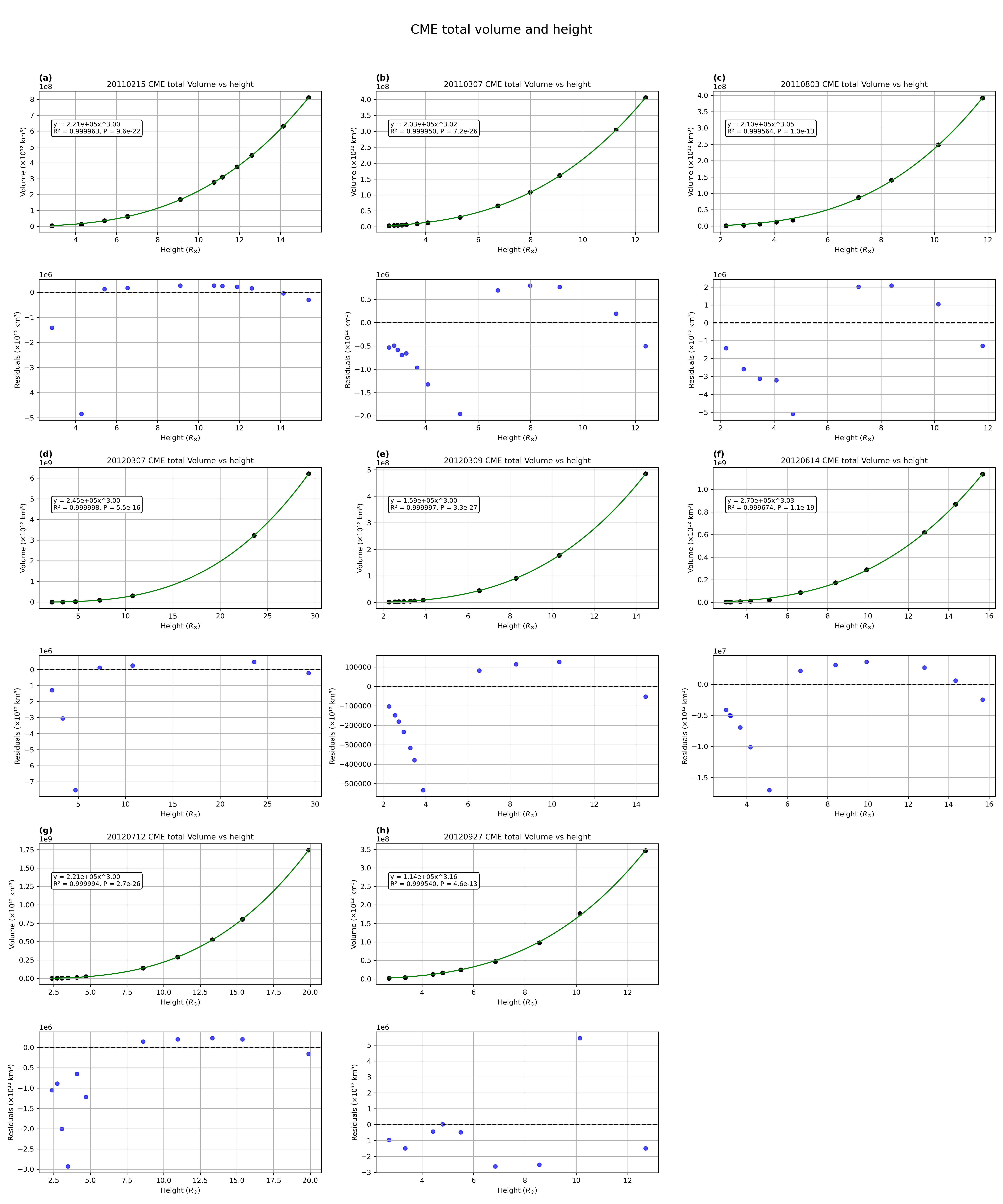}
    \caption{Power-law fits to the total CME volume ($Vol$) as a function of height for eight events. Panels 1, 3, and 5 illustrate the volume-height data points that are represented by black dots, and the power law fits are represented by the green lines. while the panels 2, 4, and 6 show the corresponding residuals.}
    \label{powerlaw_all_CMEs}
\end{figure}

Plots displaying the volume versus height (distance from the Sun) show that power-law provides an excellent description of the relationship between these parameters. 
Figures~\ref{powerlaw_20110804} and~\ref{powerlaw_20110907} illustrate the power-law fits to the volume–height data for the events on 2011 August 4 and 2011 September 7, respectively.  In the successive panels (upper row) we present the considered dependence separately for the three components of flux rope's volume and finally for its entire volume. The residuals of the applied fit are displayed in the bottom row to confirm the accuracy of the fitting procedure with the power-law function. The coefficient of determination ($R^2$) value of 0.99 indicates that the fitted model explains approximately 99\% of the variance in the volume, while the low $p$-value confirms the statistical significance of the fit.
The same plots for the remaining eight CMEs are displayed in Figure~\ref{powerlaw_all_CMEs}. However, in this case, the results are only shown for the total volume of CMEs.

By examining the individual plots, in particular the residuals distribution, one can observe that the individual volume components (the ellipsoidal front, the asymmetric disk, and the conical legs) of the CMEs evolve at different rates. This implies that the volume of the CME does not expand uniformly as it moves from the inner to the outer corona. The residual distributions show that almost all CMEs in their early stages of expansion, up to about\textbf{ $4R_{\odot}-7R_{\odot}$,} have negative residuals. This indicates that the power-law, which was fitted to all data points, underestimates the expansion. The trend changes clearly as we move farther away. The residuals start to take on positive values, which means that the expansion is overestimated by the used power-law function. This implies that the CME expansion in the inner and outer corona undergoes two distinct evolutionary phases. These results are consistent with the findings of \citet{Majumdar_2022}.
We couldn't directly confirm the two different power-law patterns of CME volume evolution in the inner and outer corona because we didn't get enough data points in the inner corona.
Additional observations could potentially be obtained by utilizing the EUVI instrument from STEREO-A and STEREO-B (see e.g., \citealp{Gopalswamy_2018}). However, the relationship between the white-light features observed by coronagraphs and extreme-UV structures is not straightforward. During the transition of the CMEs from EUVI to coronagraph FOV, mismatches could introduce uncertainties, as the extreme-UV emission tracks plasma at specific temperatures and not necessarily the full extent of the flux rope. Because of this, we have opted not to incorporate reconstructions based on EUVI into this analysis.

The larger power-law indices of  the ellipsoidal front and asymmetric disk suggests that the flux rope structure exhibits a differential expansion pattern, as the leading parts of the CME expand at a faster rate than the legs. A similar difference can also be noted in the case of residuals. Figure \ref{powerlaw_20110907} clearly demonstrates that the residuals in the case of legs exhibit the opposite trend from those in the other flux-rope elements and the entire volume. Legs residuals show positive values in the inner corona, suggesting that the power-law applied overestimates expansion. The situation is reversed in the outer corona. 

\begin{figure}[htbp]
    \centering
    \includegraphics[width=0.5\textwidth]{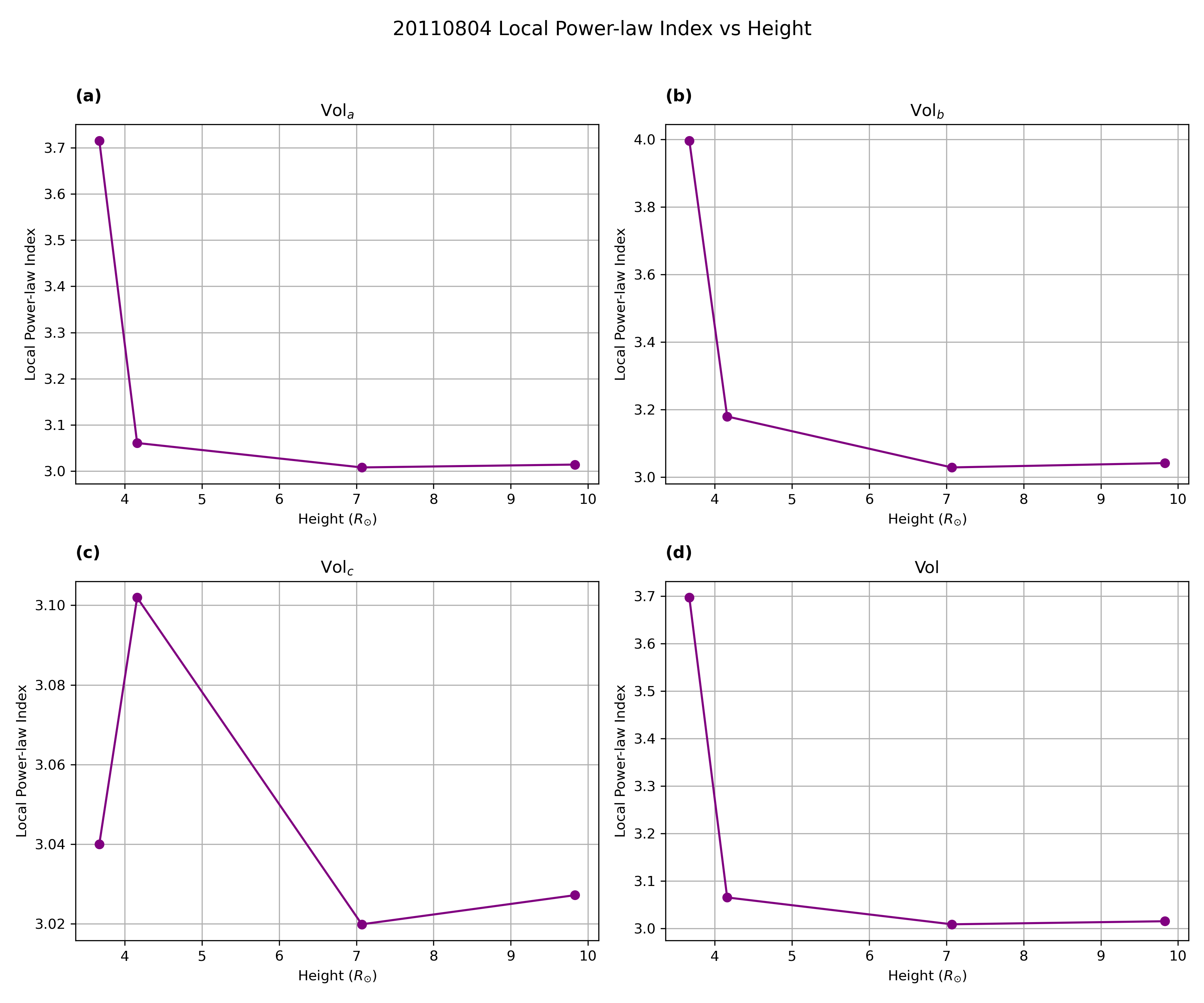}
    \caption{The local power-law index of each volume component (e.g., $Vol_a, Vol_b, Vol_c)$ and the total volume, $Vol,$ as a function of height for the CME on 2011 August 4. }
    \label{index_powerlaw_20110804}
\end{figure}

\begin{figure}[htbp]
    \centering
    \includegraphics[width=0.5\textwidth]{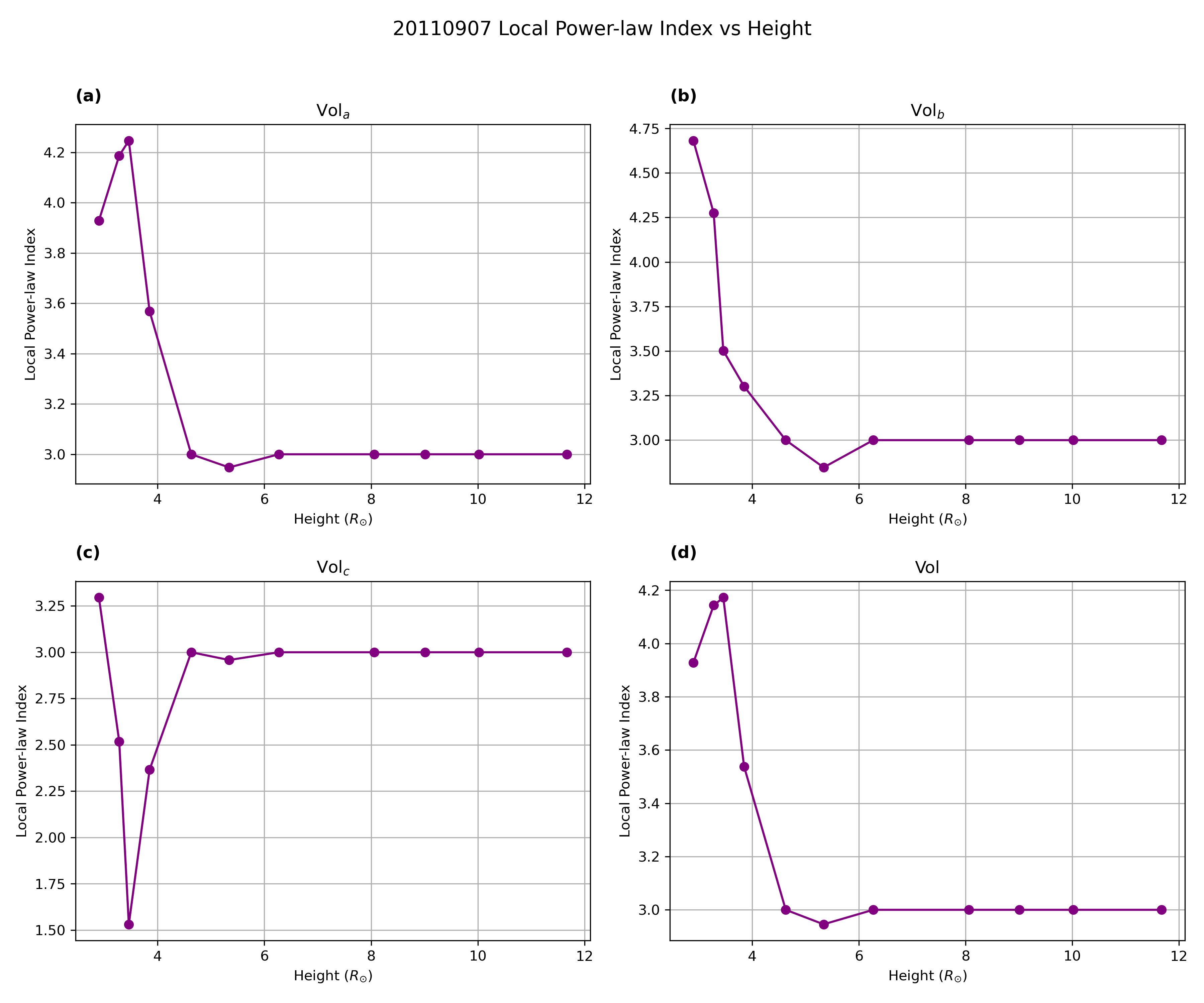}
    \caption{The local power-law index of each volume component (e.g., $Vol_a, Vol_b, Vol_c)$ and the total volume, $Vol,$ as a function of height for the CME on 2011 September 7. }
    \label{index_powerlaw_20110907}
\end{figure}

Figures ~\ref{index_powerlaw_20110804} and~\ref{index_powerlaw_20110907} show the local power-law index of each volume component (e.g., $Vol_a, Vol_b, Vol_c, and$ $Vol$) as a function of
height for the CMEs on 7 September 2011 and 4 August 2011. The local power law index is determined using the following equation:\begin{equation}\centering\text{local power law index} = \frac{\log(\text{CME volume})}{\log(\text{leading edge height})}\end{equation}

We observe that the local power-law index of the CME volume expansion reaches a maximum in the inner corona, subsequently decreases to a minimum, and ultimately saturates at larger distances from the Sun.  This trend is consistent with the model that CMEs overexpand at first, then gradually decrease in volumetric growth rate to enter a phase of steady expansion at larger heliocentric distances. This finding is also in accordance with the physical processes that are responsible for the expansion of CMEs. The ejections are accelerated by the Lorentz force, which gradually decreases as the distance of the events from the Sun increases. Evolution of CME width with height show a similar trend \citep{2022ApJ...936..122D}.
We have observed similar trends in the rest of the events considered in our study.

Since the GCS model is a purely geometric construction, the volumes of the front, disk, and legs are mathematically linked. Therefore it is expected their volumetric evolution to exhibit similar scaling behavior with height. However, our observations show that their local power-law indices differ significantly at lower coronal heights, indicating the individual components of the CME flux rope expand at different rates.
At larger heliocentric distances, however, the local power-law indices of the front, disk, and legs converge toward a value of $\sim 3$ (Figures ~\ref{index_powerlaw_20110804} and~\ref{index_powerlaw_20110907}), indicating the onset of self-similar expansion.

\subsection{Association between flare and CME kinematics}
\label{flare-cmekinematics}
\begin{figure}[htbp]
    \centering
    \includegraphics[width=0.5\textwidth]{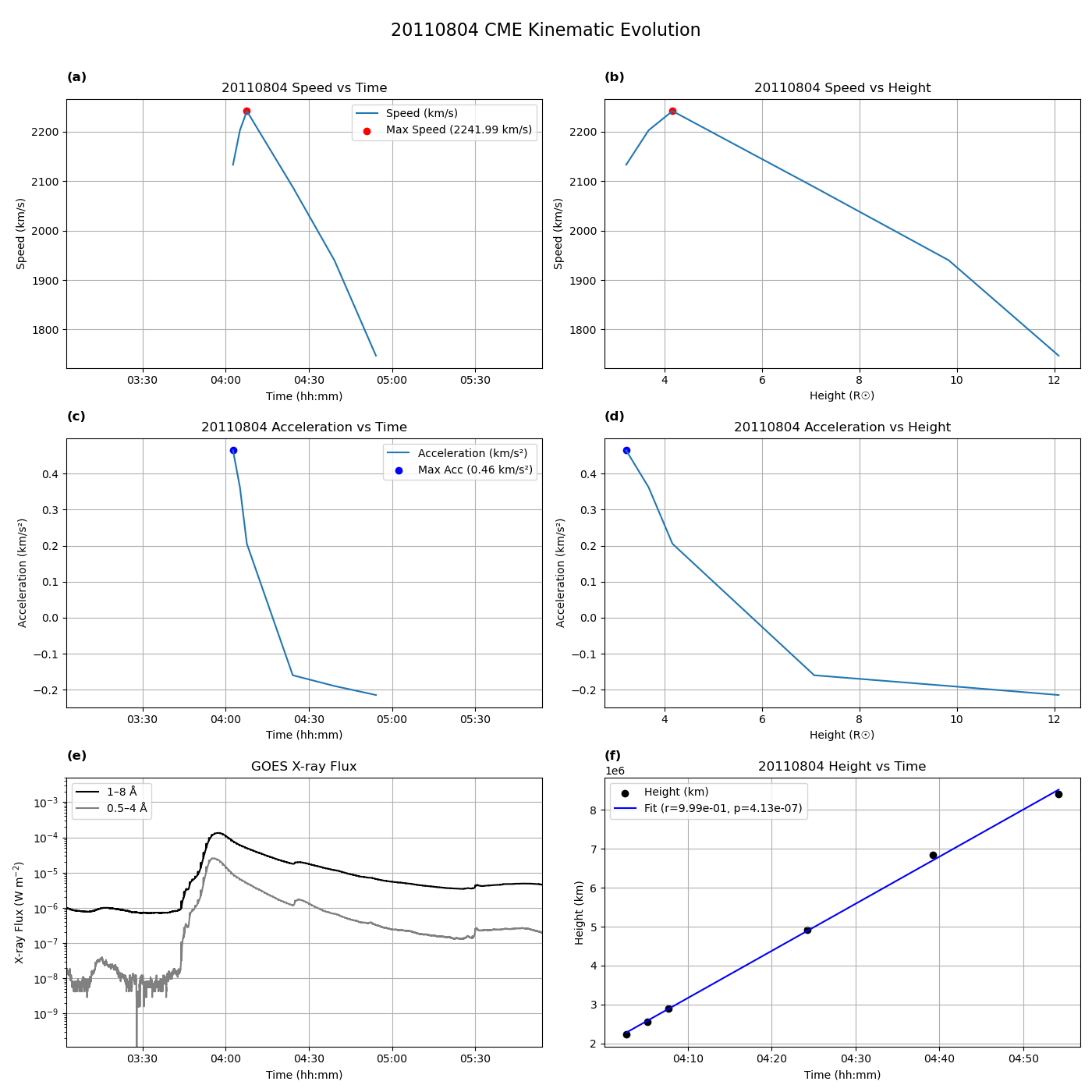}
    \caption{Kinematic evolution profile of the CME on 2011 August 4 and its correspondence with the X-ray flux of the associated solar flare. Top row: CME speed versus time and height. Middle row: acceleration versus time and height. Bottom panels show GOES X-ray flux and the linear fit to CME height-time data points.}
    \label{kinematics_20110804}
\end{figure}

\begin{figure}[htbp]
    \centering
    \includegraphics[width=0.5\textwidth]{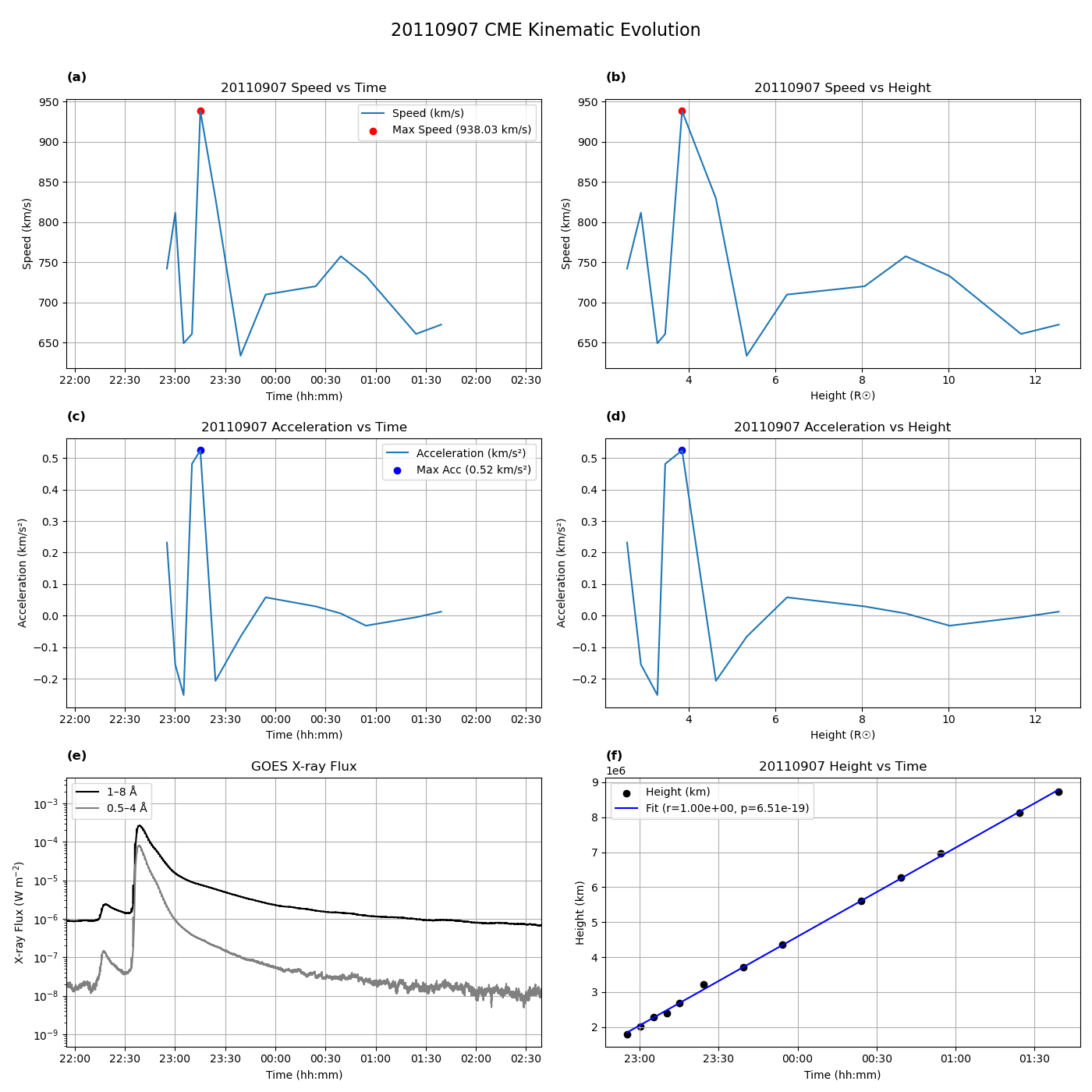}
    \caption{Kinematic evolution profile of the CME on 2011 September 7. Top row: CME speed versus time and height. Middle row: acceleration versus time and height. Bottom panels show GOES X-ray flux and the linear fit to CME height-time data points.}
    \label{kinematics_20110907}
\end{figure}

It is well established that the CME kinematics have a very strong relation to the associated flare X-ray flux (\citealp{Harrison1995,Zhang_2001,https://doi.org/10.1029/2005JA011151,Temmer_2008}). The impulsive acceleration phase of such CMEs is often well aligned with the rising phase of the flare X-ray flux. The high impulsive acceleration of some of the studied events have been reported in \citet{Gopalswamy2021CommonOrigin}.
Figures~\ref{kinematics_20110804} and~\ref{kinematics_20110907} illustrate the strong temporal association between the CME kinematics and the associated flare flux, as we observe the maximum CME speed is achieved within ~30 min of the peak of the associated flare. In the case of the faster CME that occurred on 2011 August 4, the maximum speed of approximately $2241~\mathrm{km~s^{-1}}$ was reached at a height of $4.16~R_{\odot}$. Beyond this point, the CME exhibited a consistent deceleration throughout the FOV of our observations. This behavior is typical for fast CMEs, where drag forces from the ambient solar wind become dominant, resulting in a sharp deceleration.

In contrast, the slower CME from 2011 September 9 appeared to approach a near-zero acceleration state within the observed FOV. This event reached a maximum speed of $938~\mathrm{km~s^{-1}}$ at a height of $3.85~R_{\odot}$. In both CMEs, the peak acceleration occurred within approximately 30 minutes of the peak in the GOES soft X-ray flux. The CME velocity peak was attained afterwards. Such a strong temporal correlation between the evolution of the CME kinematics and the GOES soft X-ray flux of the associated flare suggests likely causal connection between the flare and CME kinematics.

\subsection{Association between flare and CME volumetric evolution}
\label{flare-cmevolume}
\begin{figure}[htbp]
    \centering
    \includegraphics[width=\textwidth]{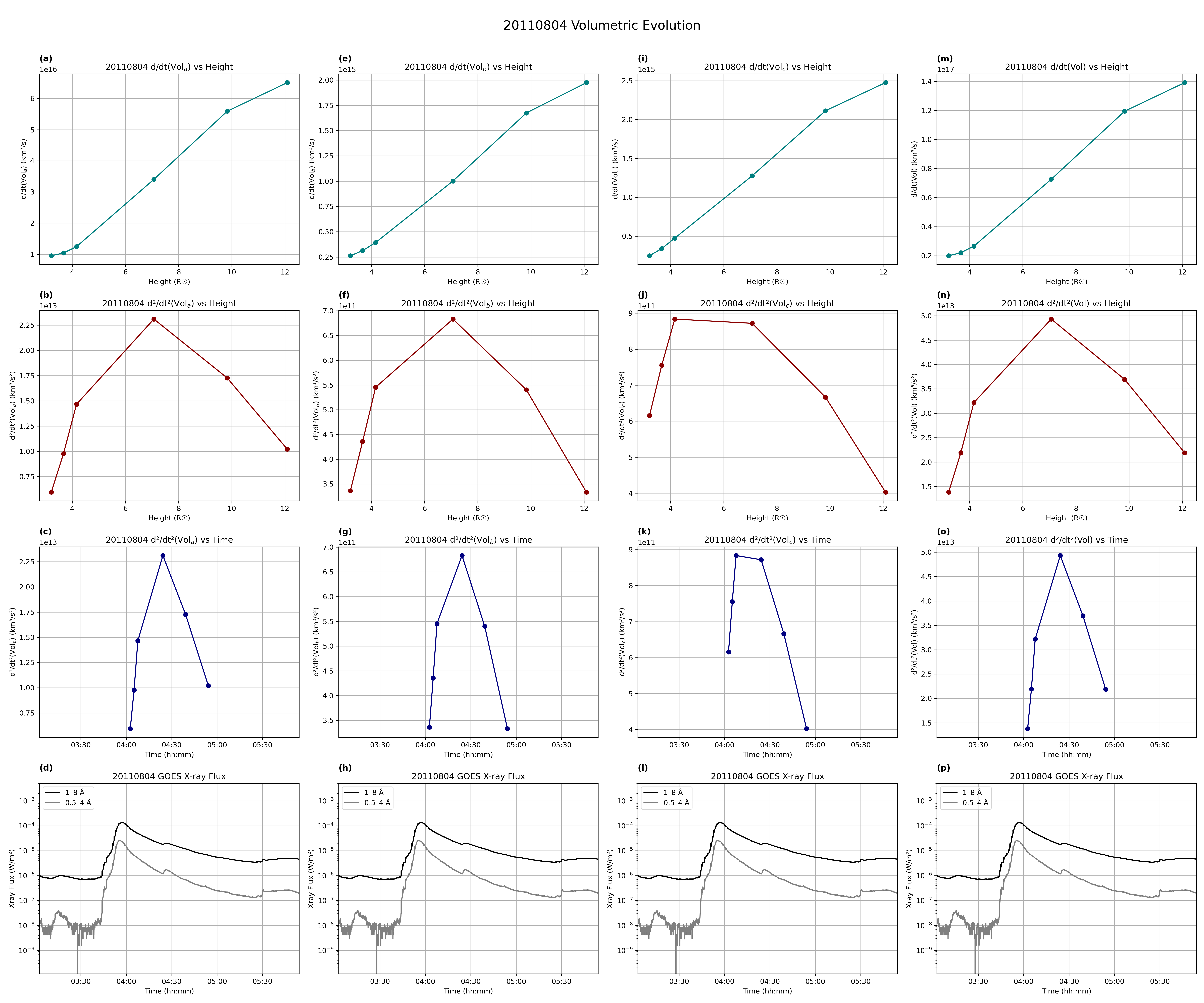}
    \caption{Volumetric evolution of the 2011 August 4 CME and its correspondence with the X-ray flux of the associated solar flare. The first and second rows contain the first and second order time derivatives of each volume component (e.g., $Vol_a, Vol_b, Vol_c)$ and the total volume, $Vol,$ respectively, and are plotted against CME leading edge height. The third row shows the second-order temporal derivative of each volume component plotted against time. Finally, in the fourth row, GOES X-ray flux is plotted.}
    \label{volume_20110804}
\end{figure}

\begin{figure}[htbp]
    \centering
    \includegraphics[width=\textwidth]{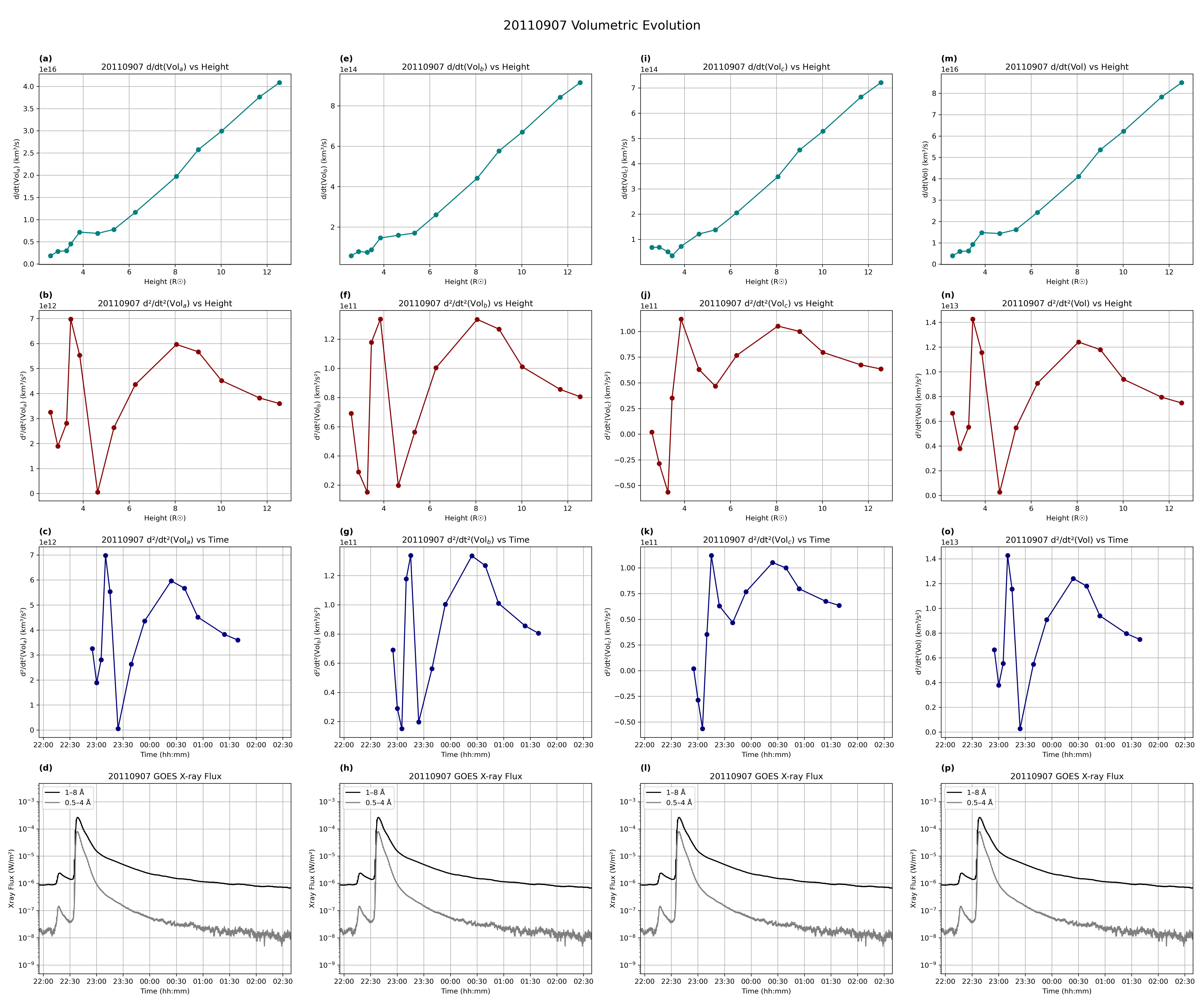}
    \caption{Volumetric evolution of the 2011 September 7 CME and its correspondence with the X-ray flux of the associated solar flare. The first and second rows contain the first and second order time derivatives of each volume component (e.g., $Vol_a, Vol_b, Vol_c)$ and the total volume, $Vol,$ respectively, and are plotted against CME leading edge height. The third row shows the second-order temporal derivative of each volume component plotted against time. Finally, in the fourth row, GOES X-ray flux is plotted.}
    \label{volume_20110907}
\end{figure}

Figures~\ref{volume_20110804} and~\ref{volume_20110907} demonstrate that the volume expansion rate $Vol'$ generally increases with heliocentric distance. However, different structural components of the CME flux rope exhibit distinct expansion characteristics.

In the case of the faster CME shown in Figure~\ref{volume_20110804}, the second-order derivative of volume $Vol''$ reveals an impulsive phase. We observe a rapid increase in $Vol''$ that peaks at approximately $7.07~R_{\odot}$, which coincides with the rise in GOES soft X-ray flux.  $Vol''$ subsequently decreases sharply and has not yet reached saturation within the observed FOV, consistent with the kinematic behavior of this CME.

Conversely, in the slower event illustrated in Figure~\ref{volume_20110907}, $Vol''$ reaches a maximum at $3.46~R_{\odot}$, followed by a steep decline and then a gradual rise toward a saturated value. This multi-phase behavior in the volumetric evolution is consistent with the corresponding kinematic evolution and the flare X-ray flux.
\begin{figure}[htbp]
    \centering
    \includegraphics[width=\textwidth]{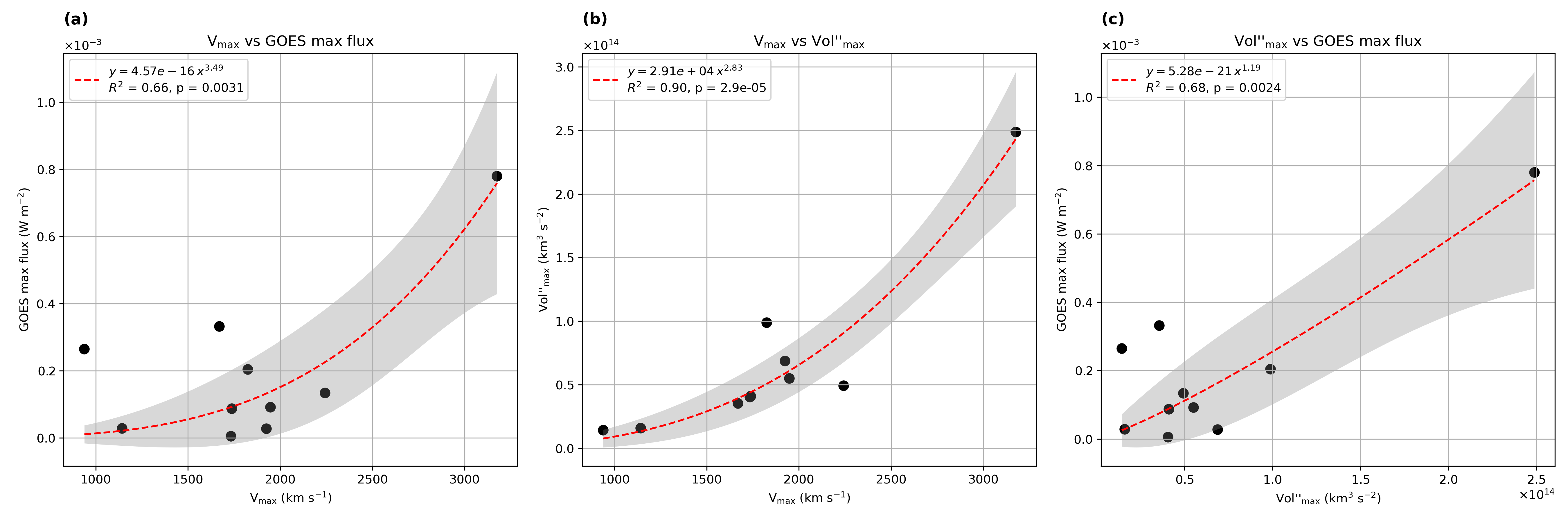}
    \caption{Panel (a): Relationship between the maximum speed of CMEs ($V_{\rm max}$) and the peak GOES X-ray flux of the associated solar flares. Panel (b): Relationship between $V_{\rm max}$ and  $Vol''_{\rm max}$ of CMEs. Panel (c): Relation between $Vol''_{\rm max}$ and the flare's peak X-ray flux. The shaded region in all of the panels represent uncertainty in the model fit.}
    \label{cme_flare_kinematics_volume}
\end{figure}

The Figure \ref{cme_flare_kinematics_volume} shows the relation between the flare X-ray flux with the kinematic and volumetric evolution of the CMEs.
In panel~\ref{cme_flare_kinematics_volume}(a), we see a strong correlation between the maximum speed of CMEs ($V_{\rm max}$) and the maximum soft X-ray flux of the associated solar flares. The power law, which fit best these data points, suggests that higher flare flux corresponds to greater CME speeds. This trend is attributed to the reconnection flux that is common to the flare and CME in an eruptive event (\citealp{GOPALSWAMY2018}). More intriguingly, panels~\ref{cme_flare_kinematics_volume}(b) and~\ref{cme_flare_kinematics_volume}(c) reveal a similarly strong correlation between the maximum  $Vol''$ and both the flare's peak X-ray flux and $V_{\rm max}$. These findings suggest that energy release in the eruption not only influences CME kinematics but also directly contributes to the expansion dynamics of the CME. The strong coupling of maximum  $Vol''$ with both flare and $V_{\rm max}$ points to a potentially important role of magnetic reconnection in driving volumetric expansion during CME evolution.

To our knowledge, this is the first time a temporal association of multi-phase volumetric evolution, with both the kinematic profiles and the X-ray flux of the associated flare, has been reported. The observed strong correlation of $Vol''$ with both the CMEs’ kinematics and the flare X-ray flux suggests a likely causal connection between flare energy release and the volumetric as well as kinematic evolution of CMEs.

\section{Summary and Conclusion}
\label{sec:conclusion}

In this study, we investigated the volumetric and kinematic evolution of ten Earth-directed CMEs, all of which were associated with solar flares, exhibited no interaction with other CMEs, and originated within $\pm$45$^{\circ}$ of the solar disc center. We employed multi-viewpoint observations from COR1 and COR2 onboard STEREO-A and STEREO-B, along with LASCO C2 onboard SOHO to perform GCS reconstructions of the CMEs. The three vantage point observations allowed us to better constrain key geometrical parameters such as the aspect ratio, half-angle, and leading-edge height, which are essential inputs for computing CME flux rope volumes using the method developed by \citet{Holzknecht}.

For all events analyzed, the relationship between CME volume and leading edge height was best described by a power-law fit. Interestingly, the volumetric expansion was found to be non-uniform across the CME structure; different components of the flux rope exhibited distinct power-law behaviors. The leading regions (e.g., the ellipsoidal front and asymmetric disk) expanded more rapidly than the conical legs.
These findings are consistent with those of \citet{Majumdar_2022}. They have also reported that CMEs follow two separate power-law trends in the inner and outer corona, with a transition near $4~R_{\odot}$.  The coronal magnetic field is highly dynamic and non-homogeneous, and is strongly influenced by large-scale structures such as streamers and coronal holes. Determining a precise height that demarcates the boundary between the inner and outer corona would require a more robust analysis involving a larger sample of events and observations at lower coronal heights with higher cadence. With such an analysis, we anticipate that different CMEs, depending on their position angle, will experience a different boundary of inner and outer corona.
However, an analysis of the local power-law index of CME volumetric expansion revealed a multi-phase pattern. It showed a peak at lower coronal heights ($4~R_{\odot}$- $6~R_{\odot}$), followed by a gradual decline, and eventually it saturated at larger heliocentric distances .  This suggests that CMEs undergo an initial over-expansion phase, after which their volumetric growth slows and stabilizes as the CME propagates further.  

The primary purpose of this work was to study the volumetric and kinematic growth of the CMEs and examine if they exhibit similar patterns in their evolution, also to explore if the volumetric evolution shows any association with the flare x-ray flux. The CME speed and acceleration profiles displayed the characteristic three-phase evolution: an impulsive acceleration phase, a gradual deceleration phase, and a final regime of near-zero acceleration. All CMEs reached maximum speeds between $3$–$5~R_{\odot}$, consistent with previous studies \citep{Sachdeva2017, Ravishankar2019, Pattnaik_2025}, as well as a strong temporal correlation was observed between the CME kinematics and the flare soft X-ray emission.

We also analyzed the second-order derivative of CME volume with time, $Vol''$, which exhibited a clear temporal correlation with both the kinematic profiles and the flare X-ray flux. In all events, the rising phase of the flare coincided with an impulsive increase in $Vol''$, which reached its peak between $7$–$10~R_{\odot}$. Thereafter, $Vol''$ went through a declining phase and finally saturated. However, in case of faster events in our list, no such saturation was seen. Suggesting that for the faster events, the point of saturation of $Vol''$ might have lied beyond our observational FOV.

It is important to note that with higher cadence data and improved FOV coverage extending further into the lower corona, the estimates of maximum $Vol''$, as well as peak speed and acceleration, could be further refined. Such improvements would provide better insight into the early expansion dynamics of CMEs. Nonetheless, the strong observed temporal correlation between CME volumetric and kinematic evolution and the associated flare activity strongly motivates us to investigate the influence of magnetic reconnection on the CME evolution.
While the present study is limited to a relatively small set of events, expanding on this work with a larger sample size and performing a robust statistical analysis would allow us to investigate important questions such as:
\begin{itemize}
    \item Whether the volumetric expansion of slow CMEs differs systematically from that of fast CME
    \item How the ambient interplanetary medium and solar cycle conditions influence CME expansion rates.
\end{itemize}

Such investigations could contribute to improving the physical understanding of CME evolution and refining space weather forecasting models.

\section{Acknowledgement}
The work done by Ashutosh Pattnaik and Grzegorz Michalek was supported by NCN through the grant 2023/49/B/ST9/00142.
We thank all the members of the STEREO/SECCHI and SOHO/LASCO consortium who built the instruments and provided the data used in this study. LASCO images are courtesy of the SOHO consortium. This CME catalog is generated and maintained at the CDAW Data Center by NASA and The Catholic University of America in cooperation with the Naval Research Laboratory. SOHO is a project of international cooperation between ESA and NASA. We acknowledge the Community Coordinated Modeling Center (CCMC) at Goddard Space Flight Center for the use of the \href{https://iswa.gsfc.nasa.gov/IswaSystemWebApp/}{Integrated Space Weather Analysis(ISWA) system}. 
\bibliography{bibliography.bib}

\begin{appendix}

\begin{table}[htbp]
\centering
\caption{Flare-associated CMEs' source region information}
\label{CME_source_region}
\begin{tabular}{llll}
\hline
\textbf{Flare Onset} & \textbf{Flare Location} & \textbf{Flare Class} & \textbf{CME Onset} \\
\hline
2011-02-15 01:44 & S20W10 & X2.2  & 02:25 \\
2011-03-07 13:44 & N11E21 & M2.0  & 14:40 \\
2011-08-03 13:17 & N17W29 & M6.0  & 13:55 \\
2011-08-04 03:41 & N15W39 & M9.3  & 04:10 \\
2011-09-07 22:32 & N15W31 & X1.8  & 23:24 \\
2012-03-07 00:02 & N18E31 & X5.4  & 00:36 \\
2012-03-09 03:22& N17E02 & M6.3  & 04:25 \\
2012-06-14 12:52 & S17W00 & M1.9  & 14:09 \\
2012-07-12 15:37 & S14W02 & X1.4  & 16:54 \\
2012-09-27 23:36 & N09W26 & C3.7  & 00:12 \\
\hline
\end{tabular}
\end{table}

\end{appendix}
\end{document}